\documentstyle[11pt,aaspp4]{article}
\begin{document}
\baselineskip=25pt
\newcommand{\be}{\begin{equation}}
\newcommand{\ee}{\end{equation}}
\title{Clusters of galaxies as a storage room for cosmic rays}
\author{V.S. Berezinsky\altaffilmark{1}, P. Blasi\altaffilmark{1,2} 
and V.S.Ptuskin\altaffilmark{3}}
\altaffiltext{1}{Laboratori Nazionali del Gran Sasso\\
Statale 17 bis 67010 - Assergi (L'Aquila) - ITALY}\par
\altaffiltext{2}{Dipartimento di Fisica, Universit\'a degli Studi di L'Aquila\\
Via Vetoio, 67100 Coppito (L'Aquila) - ITALY}\par
\altaffiltext{3}{Institute of Terrestrial Magnetism, Ionosphere 
and Propagation of Radiowaves\\
Troitsk, Moscow region - 142092, Russia}\par
\vskip 1cm
\begin{center}
{\it To be published in The Astrophysical Journal}
\end{center}
\vskip 1cm
\begin{abstract}
It is demonstrated that clusters of galaxies are able to keep cosmic rays
for a time exceeding the age of the Universe. This phenomenon reveals 
itself by the production of the diffuse flux of high energy gamma and neutrino
radiation due to the interaction of the cosmic rays with the intracluster gas. 
The produced flux is determined by the 
cosmological density of baryons, $\Omega_b$, if a large
part of  this density is 
provided by the intracluster gas. The signal from relic cosmic rays has
to be compared with the flux produced by the late sources, which can be 
considered as a background in the search for cosmic ray production in
the past. We calculate this flux considering the normal 
galaxies and AGN in the clusters as the sources of cosmic rays. Another 
potential cosmic ray source is the shock in the gas accreting to a cluster. 
We found that this background is relatively high: the diffuse fluxes 
produced by  relic cosmic rays are of the same order of magnitude which can 
be expected from AGN in the clusters. In all cases the predicted diffuse
gamma-ray flux is  smaller than the observed one and the diffuse neutrino
flux can be seen as the small bump at $E\sim 10^6~GeV$ over the atmospheric 
neutrino flux. A bright phase in the galaxy evolution can be a source of the 
relic cosmic rays in clusters, revealing itself by
diffuse gamma and neutrino radiations. We found that the observation 
of a signal from the bright phase is better for an individual cluster.
\end{abstract}

 
\section {Introduction}
 The observations of X-ray radiation 
from the clusters of galaxies evidence that a large fraction of the 
 cosmological baryonic density, $\rho_b^{cos}$ is given by the 
intracluster gas (White and Frenk 1991, White 1993, White and Fabian 1995)
\markcite{WhFr}\markcite{Wh}\markcite{WhFa}. 
More specifically the cosmological density
provided by the mass of the intracluster gas, $\Omega_b^{cl}$ is close to 
that derived from nucleosynthesis $\Omega_b^{ns}$. If $\Omega_0=1$ one 
finds from the references above $\Omega_b^{cl}h^{3/2} \approx 0.05$
to be compared with the nucleosynthesis value 
$0.009\leq \Omega_b^{ns} h^2 \leq 0.024$ 
(Copi, Schramm and Turner 1995)\markcite{CShT}.

This observation was recently used by Dar and Shaviv (Dar and Shaviv 1995a, 
Dar and Shaviv 1995b) \markcite{DSprl,DSaph} for calculating 
high energy gamma and neutrino radiations from 
clusters of galaxies (see references therein for the earlier works). 
In contrast to their calculations, where 
some ad hoc assumptions were made (most notably one about universality 
of cosmic 
ray flux in the galaxies and clusters), we shall use here a standard picture of 
production and propagation of cosmic rays (CR). This approach results in a
lower flux 
and a different spectral index of radiation at high energy as compared with 
the calculations by Dar and Shaviv. Our conclusion about this 
low flux is valid for 
all present-day sources of cosmic rays considered in the literature, 
namely, normal galaxies and AGN in clusters, intracluster and accretion  shocks.

However, as we shall show here, the clusters of galaxies are able to keep the 
accelerated particles for cosmologically long times and thus the 
production of gamma and neutrino radiations at present epochs can be due
to particles accelerated in the past. 
The effective confinement of cosmic rays in the 
clusters was recently recognized by Volk, Aharonian and Breitschwerdt (1996) 
\markcite{Volk}. They considered 
the cosmic rays in the clusters from the starburst galaxies in the past and 
TeV gamma radiation produced at present.   

A powerful source of CR production can be the bright phase in the evolution of 
galaxies. The idea of bright phase was first put forward by Partridge and 
Peebles (1967) \markcite{PaPe} and was 
further developed in many works (see e.g. (Schwartz, Ostriker and Yahil 1975,
Ostriker and Cowie 1981, Thompson 1989, Ozernoy 
and Chrnomordik 1976)\markcite{SOY}\markcite{OsC}\markcite{Th89}
\markcite{OzCh}).
The bright phase is basically connected with the fast evolution of the massive
stars (Zeldovich and Novikov 1971, Tinsley 1980)\markcite{ZeNo}\markcite{tin}. 
This process should result in  explosions and 
shocks in young galaxies and clusters, and thus in acceleration of CR. 
One particular 
mechanism, convenient for calculations, is SN explosions. This was studied 
by Schwartz, Ostriker and Yahil (1975), Ostriker and 
Cowie (1981)\markcite{SOY}\markcite{OsC}) and 
most recently, in connection with cluster of 
galaxies, by Volk et al. (1996)\markcite{Volk}.  

Clusters of galaxies are suitable for observations of bright phase in case 
it occures late, at redshifts $z \sim 1-2$ when the clusters were formed.
These observations can include a search for 
young galaxies due to blue-enhanced 
radiation and in particular due to gamma and neutrino radiation produced 
by CR in clusters.

The plan of our paper is the following. In Section 2 we discuss the propagation
 of cosmic rays and their maximum energy. 
In Section 3 the confusing problem of the 
spectrum of produced radiation will be considered. The calculations of 
diffuse fluxes of gamma and neutrino radiation due to normal galaxies are 
presented in Section 4. In Section 5 we discuss the other sources of cosmic 
rays in a cluster, namely accretion shocks and AGN. Finally, in Section 6 
we shall consider the bright phase of galaxy evolution relevant to radiation   
from clusters.

We shall finish this Introduction by a short summary of the  
properties  of galaxy clusters relevant for our calculations.

A fundamental value which characterizes a cluster is the richness, or number 
of galaxies, $N_g$, in a group or cluster. This value typically varies from 
10 to several hundreds. The clusters with $N_g > 100$ are defined as rich ones.

The total mass of a cluster, determined from the virial theorem, is called 
{\em gravitational} mass, $M_{grav}$. The cluster gravitational-mass 
distribution is given, according to Bahcall and Cen (1993)\markcite{BaCe}, 
as 
\be
n(>M)= 4\times 10^{-5} ({M\over M_*})^{-1} exp (-{M\over M_*})~h^3~Mpc^{-3}
\label{eq:mf}
\ee
where $M$ is the total (gravitational) mass of a cluster within 
$1.5 h^{-1}~Mpc$ and 
$M_*=(1.8 \pm 0.3)\times 10^{14}h^{-1}M_{\odot}$. Further on we shall refer to
the cluster with mass $M_*$ as the {\it representative} cluster.
The fraction of the gas 
mass in a cluster (relative to the gravitational mass) is commonly estimated as 
$(0.05 - 0.1)h^{-1.5}$ (see (B\"{o}hringer 1995, Bahcall 1995)
\markcite{Bor, Bahc}).
In a recent work (White and Fabian 1995)\markcite{WhFa} 
this fraction is found for a sample of 13 clusters to be 
between 10 and 22 per cent.

The distribution (\ref{eq:mf}) covers 
the range from $M_{grav} \sim 10^{12} h^{-1}M_{\odot}$ (small groups) to 
$5\times 10^{15}h^{-1} M_{\odot}$. The representative mass of a cluster, 
according to distribution (\ref{eq:mf}) is 
$M_*=1.8 \times 10^{14} h^{-1} M_{\odot}$. The space density of clusters 
with $M > M_*$ is $\sim 1.4\times 10^{-5}~Mpc^{-3}$.

The size of a cluster, $R_{cl}$, is defined as the distance at which 
a galaxy is gravitationally confined. Thus, this size can be determined by the
velocity dispersion of galaxies $\sigma \approx (<v_{gal}^2>)^{1/2}$ as 
\be
R_{cl} = G M_{grav}/<v_{gal}^2>
\label{eq:vir-gal}
\ee
and is typically $1.5h^{-1}~Mpc$, or roughly between $2 - 3~Mpc$.

The hot gas is confined to a cluster due to gravitational potential of the 
system. The virial theorem implies the equilibrium temperature of the gas
\be
3kT_{eq}=m_p<v_{gas}^2>=G M_{grav}m_p/R_{cl}
\label{eq:vir-gas}
\ee
The galaxies in a cluster move in the same gravitational potential and thus 
from Eq's(\ref{eq:vir-gal}) and (\ref{eq:vir-gas}) one has
\be
T_{eq} \sim m_p \sigma/(3k) \sim 4\times 10^7 
\left( \frac{\sigma}{10^8~cm/s} \right) ^2~ K
\ee
If the temperature of intracluster gas were much larger than $T_{eq}$,
the gas expands, if much lower - - it condenses.

Clusters of galaxies are powerful sources of X-ray radiation 
($L_X \sim 10^{42} - 10^{44}~erg/s$). This radiation is interpreted as 
bremsstrahlung radiation of hot intracluster gas with temperature 
$T \sim (1 - 10)\times 10^7~K$, which is indeed close to the equilibrium 
temperature.  

Both spiral and elliptical galaxies are observed in the clusters. The 
spiral rich clusters have more than $50\%$ of spirals, while the spiral 
poor clusters - - only $30\%$ (Gorbatskii 1988)\markcite{Go} 
(see also (Dressler 1980)\markcite{Dr}). However, many observers now 
think that elliptical galaxies are more abundant in rich clusters 
(Colafrancesco,  private communication).\par
More than $10 - 20\%$ of clusters have in the center a 
bright powerful $cD$ galaxy 
with an absolute magnitude $M_V < - 20^m$. This galaxy usually provides 
a large part of radio emission of a cluster. Other radiogalaxies are 
also observed in clusters. Thus, one can assume the presence of AGN's 
in clusters. 

\section{Propagation of cosmic rays and maximum energy}

In the sections 2-4 we shall limit ourselves only to cosmic rays produced 
by normal galaxies in a cluster. This gives a conservative estimate of the 
produced flux, which can be considered as the lower limit, and 
also gives {\em reference calculations}: using them we shall discuss 
possible contributions of other CR sources.

Our consideration is based on two main observations.
\begin{itemize}
\item
The diffuse flux is directly connected with the cosmological density of 
baryons and not with the gas density in a cluster.
\item
The spectrum of radiation is determined by the {\em production spectrum} 
of CR in the sources and not by the equlibrium spectrum of CR in a cluster.
\end{itemize}

As a result one can notice that the problem we study, is defined by a very 
restricted set of parameters, namely, by the cosmological parameter $\Omega_b$,
by the cluster richness $N_g$ (number of galaxies), by the size 
and age of clusters
$R_{cl}$ and $\tau_{cl}$, respectively, and by the generation function of CR 
in a typical galaxy, $Q_{CR}^{gal}(E)$.

Using CR production in our Galaxy as a guide, we can assume the generation 
spectrum in a galaxy to be
$Q_p^{gal}(E)=A(E+E_0)^{-\gamma_g}$ with $E_0 \sim 1~GeV$. For $E\gg 1~GeV$
one then obtains
\begin{equation}
Q_p^{gal}(E)=
(\gamma_g-1)(\gamma_g-2)\frac{L_p}{E_0^2} (\frac{E}{E_0})^{-\gamma_g}
\label{eq:gen}
\end{equation}
\noindent
where $L_p \sim 3\times 10^{40}~ erg/s$ and $2.1 \leq \gamma_g\leq 2.4$ 
(Berezinsky et al. 1990, chapter 3)\markcite{BBDGP}.
Note, that using our Galaxy as a reference source in a cluster results in a
somewhat overestimated diffuse flux from the clusters.
As it will be understood from the following consideration, we do not need 
the detailed picture of CR propagation in 
a cluster for the calculation of fluxes. 
It is only important for us to demonstrate that the escape time of 
CR out of a cluster is larger than the age of the cluster, i.e. that the 
particles injected by galaxies are confined within the cluster during 
its age.

In the case of normal galaxies the relevant estimates can be made 
using an assumption of diffusive propagation. However, in case of a 
powerful outburst, as in the case of the "bright phase", the galactic wind can 
play a crucial role in CR propagation.

Observations of radio emission from some clusters
evidence that relativistic electrons and magnetic field are distributed
there on the megaparsec scale (see e.g. Hanisch et al. 1987  \markcite{Han}). 
Magnetic field with a typical strength of a few $\mu G$ and a
principal length scale of $10-100~ kpc$ was detected in many clusters of 
galaxies by the Faraday rotation of radiation from radio sources 
(Kim et al. 1991\markcite{Kim} and Kronberg 1994 \markcite{Kron}).

The stochastic acceleration of electrons {\em in situ} on Alfv\'en 
waves in the 
Coma cluster can explain, according to Schlickeiser et al. (1987) 
\markcite{Schl}, the observed synchrotron radio emission on a
cluster scale if the diffusion coefficient of $GeV$ particles is 
$(1-4)\times 10^{29}~cm^2/s$ if the equipartition magnetic field is assumed
($B \approx 2~\mu G$).

Taking the diffusion coefficient $D_0=2\times 10^{29}~cm^2/s$
for particles with energy $E_0=1~GeV$ and using the dependence  
$D\propto E^{1/3}$, characteristic for particle scattering off the magnetic 
irregularities with the Kolmogorov spectrum, one can estimate the escape time 
from a cluster as
\be
\tau_{esc}
\approx \frac{R_{cl}^2}{6D(E)}=2.3\times 10^{12} \left( \frac{R_{cl}}
{3Mpc}\right)^2 \left( \frac{2\times 10^{29}cm^2/s}{D_0}\right) 
\left( \frac{E}{E_0}\right) ^{-1/3}~yr.
\label{eq:esc}
\ee

Thus, for energies $E\leq 2\times 10^7~GeV$ the escape time $\tau_{esc}$ is 
greater than the age of the cluster $\tau_{cl}$, for which we shall use here and
everywhere else the age of the Universe $t_0=0.82\times 10^{10}h_{0.8}^{-1}~yr$.
At energies higher than $E_c\sim 2\times 10^7~GeV$ the escape time becomes
less than the age of a cluster, $\tau _{cl}$, and production of
the secondaries becomes less effective. The determination of this critical
energy is sensitive to the dependence of the diffusion coefficient on energy
which is rather uncertain and depends on the spectrum of the 
random magnetic field
(see e.g. Berezinsky et al. (1990) \markcite{BBDGP}, chapter 9). In particular,
the diffusion coefficient may be almost energy independent if the turbulent 
perturbations in the intracluster medium consist of random shocks and 
discontinuities, as is typical in the case of supersonic turbulence. 
Particles with energies considerably higher than $2\times 10^7~GeV$ can be 
confined in a cluster in this case. 

The extreme case is given by the Bohm diffusion with 
$D\sim cr_H$. This is the minimum possible diffusion coefficient for 
propagation along the magnetic field lines. The confinement  condition 
$\tau_{esc} \geq \tau _{cl}$ in this case holds up to energy about $10^9~GeV$.

The effective confinement of CR in a cluster was also recognized by Volk et
al. (1996)\markcite{Volk}.

Another reason for the diminishing of the efficiency of gamma and neutrino
production at very high energies can be related to the maximum energy 
$E_{max}$ reached in the process of CR acceleration at the source. For the
diffusive shock acceleration by a SN blast wave moving in the interstellar
medium, this energy is known to be $\sim 10^5-10^6~GeV$ for protons. SN
shock acceleration in predecessor stellar winds of red giants and the
Wolf-Rayet stars gives $10^7~GeV$ (Volk and Biermann 1988)\markcite{vb}.  
Ensemble of shocks in OB associations may provide 
$E_{\max}\approx 10^8~GeV$ (Bykov et al. 1990 \markcite{Bykov}). 

Finally, from phenomenological consideration of the origin of ultra high energy
CR in our Galaxy, it is known (Berezinsky et al. 1990, chapter 4, 
\markcite{BBDGP} and Bird et al 1995 \markcite{Bird}) that particles with 
energies up to $\sim 10^9-10^{10}~GeV$ are most probably produced in the 
Galaxy, and the generation spectrum can have a power-law spectrum up to this 
energy.

In the following, we shall use the value $E_{max}\sim 1\times 10^8~GeV$,
either due to inefficiency of confinement or acceleration.

\section{Spectrum of radiation}
This Section is just a note clarifying the difference in the 
radiation spectrum for the cases $\tau_{esc}>\tau_{cl}$ and 
$\tau_{esc}<\tau_{cl}$.

Let us consider a nuclear-thin object, i.e one with column density much
less than the nuclear path-length. We assume that high energy protons 
diffuse from a central source. The generation function of the source is 
$Q_p \propto E^{-\gamma_g}$ and the diffusion coefficient 
$D(E) \propto E^{\eta}$.

The number of secondary particles $i=\gamma,\nu$ produced per unit time 
and unit volume for the power-law spectra of CR nucleons can be given 
in terms of the {\em production yields} $Y_i$ (Berezinsky et al. 1990)
\markcite{BBDGP}as
\begin{equation}
q_i(E,r)= Y_i \sigma_{pp}n_H c n_p(E,r)
\label{eq:pr}
\end{equation}
where $\sigma_{pp}=3.2\times 10^{-26}~cm^2$ is the normalizing cross section 
for 
high energy $pp$ scattering, $n_H$ is the number density of gas in the cluster 
and $n_p(E,r)$ is the number density of protons with energy $E$ at a distance 
$r$ from 
the source. This density  $n_p(E,r)$ can be found from the diffusion 
equation as
\begin{equation}
n_p(E,r)=\frac{1}{4\pi r}\frac{Q_p(E)}{D(E)} 
\label{eq:dens_p}
\end{equation}
The flux of i-secondaries from the 
whole source can be found by integrating (\ref{eq:pr}) 
over the volume
\begin{equation}
Q_i(E)= \int_0^{r_{max}} q_i(E,r)4\pi r^2 dr
\label{eq:q_i_1}
\end{equation}
For the case $\tau_{esc} > \tau_{cl}$ the integration is limited by the upper 
limit
\begin{equation}
r_{max}=\sqrt{6D(E)\tau_{cl}}
\end{equation}
and one obtains
\begin{equation}
Q_i(E)=3 Y_i Q_p(E)\sigma_{pp}n_Hc\tau_{cl} \sim E^{-\gamma_g}
\label{eq:fl}
\end{equation}
The physical meaning of this formula is straightforward: each high energy 
proton emitted by a source has the same probability of interaction, equal 
to $\sigma_{pp}n_Hc\tau_{cl}$. This result is exact and actually valid 
for any way of CR propagation, e.g. a rectilinear one. 

For the other case $\tau_{esc}<\tau_{cl}$, the integration in 
eq. (\ref{eq:q_i_1}) is limited by $r_{max}=R_{cl}$ and 
\begin{equation}
Q_i(E)=Y_iQ_p(E)\sigma_{pp}n_Hc\frac{R_{cl}^2}{2D(E)} \sim E^{-(\gamma_g+
\eta)}
\label{eq:fl_diff}
\end{equation}
The physical meaning of this formula is also evident: the probability of 
interaction depends on the escape time , $R_{cl}^2/D(E)$,
and thus on the energy of the particles.

Therefore, spectra (\ref{eq:fl}) and (\ref{eq:fl_diff}) are different.
This difference often causes confusion. In particular, 
Dar and Shaviv (1995a)\markcite{DSprl}
took the spectral index of gamma-radiation equal to that of the equilibrium 
spectrum of the protons in our galaxy (i.e. Eq.(\ref{eq:fl_diff})), while 
in fact it 
is equal to the index of the generation 
spectrum of protons (Eq.(\ref{eq:fl})).

To generalize this note, we mention that the absorption time of protons 
(or electrons), when essential, can play the role of age.

\section{Diffuse fluxes of high energy gammas and neutrinos due to normal
galaxies}
We consider $pp$-production of high energy gamma and neutrino radiation.
The number of gamma's or neutrinos produced by one cluster at time 
$t$ during the interval $dt$ can be given in terms of the  
{\em production yields} $Y_i$ (see Berezinsky et al. 1990, chapter 8
\markcite{BBDGP} and Table 1) as
\be
N_i(E,t)dt=Y_i Q_p^{gal}(E) N_g t\sigma_{pp} n_H c dt,
\label{eq:numb}
\ee
where $i=\gamma,\nu$, $Q_p^{gal}$ is given by Eq.(\ref{eq:gen}) and for 
other notations see Section 3. The values of the yields are reported in Table 1.
Note that due to the definition of the yields, $Y_i$ and 
$Q_i$ are taken at the same energy E.

Integrating over time and summing the 
contribution of all clusters one finds the diffuse flux as

\be
I_i(E)=\frac{3}{32\pi^2}Y_i\frac{\sigma_{pp}(ct_0)^2}{m_H R_{cl}^3}Q_p^{gal}(E)
N_g \xi \Omega_b \rho_{cr},
\label{eq:flux}
\ee
where we used $n_{cl}M_{gas}^{cl}=\xi\Omega_b\rho_{cr}$, with 
$\rho_{cr}=1.88\times 10^{-29} h^2 g/cm^3$ being the critical density of 
universe.

In Eq.(\ref{eq:flux}) we neglected the effects of cosmological evolution of 
the clusters. This is due to the fact that we limited ourselves here by normal
galaxies as the sources of cosmic rays. Cosmological evolution of normal 
galaxies is weak (X-ray observations) or absent (radio-observations). Besides
the maximum redshift for the clusters ($z_{max} \sim 2 - 3$) is small to 
produce the considerable effects for CR. The recent detailed calculations 
(Colafrancesco and Blasi\markcite{colbla}, in preparation) confirm 
that most of the contribution to gamma rays and neutrinos 
due to clusters comes from $z\lesssim 0.1$.
The exceptional case is given by the burst of CR production (bright phase)
which will be considered in Section 6.

From Eq's(\ref{eq:flux}) and (\ref{eq:gen}) one can see that the 
diffuse flux is 
determined by the cosmological density of baryons $\Omega_b$, the cluster 
parameters $N_g$ and $R_{cl}$ and the galactic quantity $Q_p^{gal}(E)$.

We can now compare the calculated gamma-ray flux (\ref{eq:flux}) with the one 
extracted from the EGRET observations at $10 - 50~GeV$
(Osborne et al. 1994)\markcite{Osborne}:
\be
I_{\gamma}^{obs}(E)=9.6\times 10^{-7}E_{GeV}^{-2.11}~cm^{-2}s^{-1}sr^{-1}
GeV^{-1}.
\label{eq:flux_exp}
\ee
To fit the data we choose $\gamma_g=2.11$, which is in the allowed range 
of the generation spectrum index. Using $\xi=1/2$, $N_g=100$, 
$Y_{\gamma}=0.116$ and $R_{cl}=1.5 h^{-1}~Mpc$ we obtain:

\begin{equation}
I_{\gamma}(E) =
3.4\times 10^{-10}~E^{-2.11}h_{80}\frac{\Omega_b h^2}{0.025} 
~cm^{-2} s^{-1} sr^{-1}GeV^{-1}.
\label{eq:flux_num}
\end{equation}
The calculated flux is only $\sim 3\times 10^{-4}$ of that observed 
experimentally. Notice that one cannot diminish arbitrarily the radius 
of clusters, $R_{cl}$, because the mass inside a smaller radius cannot 
provide $\Omega_b h^2=0.02$. 

This result is not surprising, because the column density CR traversed 
in the intracluster gas, $x \sim c m_H n_H t_0$ is of order $0.5 - 1~g/cm^2$, 
i.e. of the same order as in a galaxy. It means that fluxes of 
gamma-radiation produced by CR in the parent galaxy and in a cluster are 
of the same order of magnitude. It is known that normal galaxies cannot 
produce the observed diffuse flux at energies $100~MeV - 10~GeV$. 
Cluster production of neutrinos with the highest possible energies is of 
greater interest, because the galactic grammage $x_{gal}(E)$ diminishes 
as $E$ increases, while grammage in a cluster does not depend on energy up to
$E_p \sim 10^8~GeV$.  

\placetable{table_1}

The diffuse neutrino flux calculated with the help 
of Eq.(\ref{eq:flux}) is shown 
in Fig.1 together with vertical atmospheric neutrino fluxes. For the prompt 
atmospheric neutrinos we took the calculations of 
Thunman et al. 1995 \markcite{TIG} (see also references therein). For the 
calculated flux from clusters we used the same 
parameters as above and neutrino yields from Table 1. One can see that 
for the case of normal galaxies, as CR sources, the predicted neutrino flux is 
practically undetectable.  

The upper limit on the neutrino flux
shown in Fig.1 is obtained from the condition that the diffuse gamma-ray flux
observed at $10 - 50~GeV$ (\ref{eq:flux_exp}) is produced by clusters. It 
results in
$$
I_{\nu_{\mu}+\bar{\nu}_{\mu}}^{max}(E)=\frac{Y_{\nu_{\mu}+\bar{\nu}_{\mu}}}
{Y_{\gamma}} I_{\gamma}^{obs}(E).
$$

\section{Other sources: AGN's and shocks}
In this section we shall discuss the other sources of CR in the clusters, 
namely AGN's, shocks inside the clusters and the accretion shocks.

\subsection{Active Galactic Nuclei}
Since about $1\%$ of galaxies have an AGN, we can expect one AGN per 
cluster with richness $N_g=100$. In particular, roughly $1\%$ of spirals 
are Seyfert galaxies with luminosity $10^{44} - 10^{45}~erg/s$ 
(Weedman 1977)\markcite{Weedman}. 
About $5\%$ of elliptical galaxies are radiogalaxies
(Schmidt 1978)\markcite{Schmidt}. 
A very powerful source of CR's can be the cD-galaxy 
often observed in the center of clusters. Recently Ensslin et al. (1996)
\markcite{Ens} found that radiogalaxies in clusters can heat the 
intracluster gas due to injection of CR. 

There are no well elaborated mechanisms of particle acceleration in 
AGN's. When particles are accelerated diffusively by the accretion shocks 
in the vicinity of the black hole, they are dragged by the flow of gas 
onto the black hole. A plausible mechanism of acceleration is due to 
unipolar induction in the accretion disc around the black hole 
(Blandford 1976)\markcite{Blandford} or acceleration in jets 
(Biermann and Strittmatter 1987, Quenby and Lieu 1989)\markcite{bs,ql}. 
It can be expected that up to $10\%$ of the total energy 
release is transferred to accelerated particles. 

However, even assuming one AGN with total luminosity $L_{tot} \sim
10^{45}~ erg/s$, we arrive at $L_{CR} \sim 10^{44}~erg/s$, to be compared 
with the total CR luminosity of normal galaxies $N_gL_{CR}^g \sim
3\times 10^{42}~erg/s$. The gain factor, $30$, is still not enough to 
reconcile the Eq.s (\ref{eq:flux_exp}) and (\ref{eq:flux_num}). In the case of 
high energy neutrino radiation, this factor is important: the neutrino 
flux from clusters can produce a bump over the atmospheric neutrino flux 
at $E \sim 10^6~GeV$ (see Fig. 1).

\placefigure{fig_1}

\subsection{Shocks inside clusters}
These shocks can be produced by galactic winds. Even if this phenomenon 
occurs, it is energetically provided by SN induced shocks from the galaxies
. CR accelerated in a galaxy obtain a total energy which is about  
the same  order of magnitude as one carried away by the galactic wind. 
Actually the situation is even worse: as was pointed 
out by Breitschwerdt, McKenzie and Volk (1991)\markcite{bmv91} and
Breitschwerdt, McKenzie and Volk (1993)\markcite{bmv93}, the 
intracluster pressure is so 
large that the winds from the galaxies do not develop, except in the case of 
the starburst phase.
\subsection{Accretion shock}  
In this work we assume that a large fraction of  cosmologically produced 
baryons are concentrated in the clusters of galaxies. It implies 
a large rate of accretion of baryonic gas to the clusters. 
This  rate averaged over the age of Universe can be roughly 
estimated as
\begin{equation}   
<\dot{M}> \sim M_{gas}/t_0
\label{eq:mdot}
\end{equation}
where  $M_{gas}$ is the mass of the gas in a cluster. 

We imploy in this section the calculations of Bertschinger (1985)
\markcite{Bert}
for spherically symmetric accretion of a mixture of non-dissipative 
gas (dark matter with a critical density) and a dissipative (baryonic) gas.
The process of fluctuation growth in both linear and non-linear regimes is 
dominated by the non-dissipative component, while baryonic gas moves in 
the gravitational potential produced by dark matter. However, as demonstrated 
by Bertschinger (1985)\markcite{Bert} both 
components have the same density distribution
$\rho \propto r^{-9/4}$ and actually the independent solutions for dissipative
gas with adiabatic index $\gamma=5/3$, and non-dissipative gas are very similar.
The main difference is the existence of a shock in the baryonic gas. The shock 
originates because of the collision of the flow with  
the dense self-produced core.
It propagates 
outward according to $R_{sh} \sim t^{8/9}$, where $R_{sh}$ is the shock 
radius. At present, $t=t_0$, for intervals of time $\tau \ll t_0$
we can consider the shock as a stationary one.

The basic parameter in this theory is the turnaround radius $R_{ta}$.

The formation of a cluster takes place in the background of the 
expansion of the Universe.
$R_{ta}$ can be defined approximately as the distance 
where the free-fall velocity
$v_{ff}= \sqrt{2GM/r}$ is equal to the Hubble velocity $v_H=H_0 r$. As an  
order of magnitude this condition gives $R_{ta}^3 \sim GM/H_0^2$,
where $M$ is the total mass inside the turn-around radius $R_{ta}$. 
In our calculations we shall use a more accurate expression from 
(Gunn and Gott 1972, Bertschinger 1985)\markcite{gunn_gott,Bert}:
\be
R_{ta}=(8GMt_0^2/\pi^2)^{1/3}.
\label{eq:ta}
\ee

The solutions of Bertschinger are  scaling ones which are presented 
in terms of dimensionless quantities. As a dimension parameter for our case we 
can take the turnaround radius with value $R_{ta}=5~Mpc$. It gives the 
gravitational mass inside $R_{cl} = 2~Mpc$,
$$
M_{cl}^{grav}=\frac{4}{3}\pi\rho_{cr}R_{ta}^3 M(\lambda_{cl})=
3.1\times 10^{14}M_{\odot},
$$
where $\rho_{cr}$ is the critical cosmological density, 
$\lambda=r/R_{ta}$ is the dimensionless distance and $M(\lambda)$ is 
dimensionless scaling mass given by Bertschinger (1985)\markcite{Bert}.
In particular, for $\lambda_{cl}\equiv R_{cl}/R_{ta}=2/5$, 
$M(\lambda_{cl})=3.7$ (in this Section we use $h=0.75$).
According to 
Eq.(\ref{eq:mf}) this  mass corresponds to a representative cluster for our 
consideration.

The shock radius is found by Bertschinger to be $R_{sh}=0.347 R_{ta}$. 
Normalizing the density of baryonic gas by the condition 
$M_{cl}^{gas}/M_{cl}^{grav}=0.1$ one can find the gas density at
$R=R_{sh}$ as $\rho_{gas}=2.2\times 10^{-29}~g/cm^3$. The radial velocity 
of gas at the position of the shock is, according to 
Bertschinger (1985)\markcite{Bert},
$v_r=1.5 R_{ta}/t_0=8.4\times 10^7~cm/s$.

The particles are accelerated diffusively on the shock front which, as 
was mentioned above, can be considered as  stationary at the present time.
The particles are scattered off magnetic inhomogeneities upstream and 
downstream the accreting flux and dragged by it inside the 
cluster. The CR luminosity can be calculated
assuming that $10\%$ of kinetic energy of the infalling gas is 
transferred to accelerated particles: 

\be
L_{CR}=0.2 \pi \rho_{gas}(R_{sh})R_{sh}^2 v_r^3=2.3\times 10^{44}~erg/s.
\label{L_sh}
\ee

The maximum energy of accelerated particles can be estimated in the 
usual way. Acceleration is efficient up to energy at which the acceleration 
length,$D/u$ becomes comparable with the radius of the shock front $R_{sh}$ 
(see e.g. Jones and Ellison 1991 \markcite{JE}). 
Assuming the Bohm diffusion with $D \sim r_H c$ one obtains:
\begin{eqnarray}
E_{eV}^{max} \sim 300\frac{u_r(R_{sh})}{c}H R_{sh}\approx \nonumber\\
9.3\times 10^{17}(H/1\mu G)(R_{sh}/3~Mpc)(u_r/10^7~cm/s)~eV
\end{eqnarray}

The alternative solution for gas  accretion to the cluster is given by 
the pancake model (Zeldovich 1970)\markcite{Zel}. 
The shock is also produced in this model. 
The results of numerical calculations for the cluster-size  pancake
are given by Doroshkevich, Zeldovich and Sunyaev (1976)\markcite{DZS}. 
The distance to the shock in this simulation is 
about $0.6~Mpc$ for a typical size of a cluster $2~Mpc$. The gas velocity 
at the position of the shock is $1.5\times 10^7~cm/s$. The CR luminosity
can be estimated very roughly for the given parameters and geometry as 
 $\sim 10^{43}~erg/s$.  
\section{Bright Phase}
The first-generation stars were produced due to non-linear growth of the 
fluctuations in the baryonic gas. Their mass spectrum was strongly dominated
by the large masses with a fundamental scale given by the Jeans mass 
$M_J \sim 10^5 M_{\odot}$. The fragmentation of these massive protostars 
results in the formation of the initial mass function (IMF), being strongly 
enhanced by massive short-lived first-generation stars. Most of these 
massive stars finish their evolution as SNII.  The explosions and stellar 
winds enrich the interstellar gas by heavy elements and this composition 
is seen in metal deficient population II stars. The duration of the longest 
phase in the heavy-star evolution, the main sequence stage, can be 
estimated as the ratio of the thermonuclear production of energy, 
$\epsilon Mc^2$, to the luminosity $L$, or 
$$
\tau_{MS} \sim \frac{\epsilon Mc^2}{(1.3 \times 10^{38} M/M_{\odot})}
(L/L_{Edd})^{-1} \sim 1.4\times 10^6 \frac{L_{Edd}}{L}~ yr,
$$
\noindent
where the conversion coefficient $\epsilon \approx 3.3\times 10^{-3}$ is 
taken for a typical presupernova II massive star ($M\sim 30M_{\odot}$) with 
$2M_{\odot}$ carbon-oxygen core and $12M_{\odot}$ helium shell. For these stars
$L/L_{Edd} \sim 0.1 - 0.01$ and thus the typical duration of the bright phase 
is $10^7 - 10^8$ years. Partridge and Peebles (1967)\markcite{PaPe} 
in their pioneering work assumed the duration as $3 \times 10^7~yr$.

The total energy release during the bright phase is estimated from 
nuclear conversion. Partridge and Peebles (1967)\markcite{PaPe} found 
$W \sim 3\times 10^{61}~ergs$ per galaxy; Schwartz, Ostriker and Yahil 
(1975)\markcite{SOY} 
estimated it as $W \sim 10^{61} - 10^{62}~ergs$. These authors considered 
this energy as being released in the form of kinetic energy of SN shells, 
which results in the shocks and acceleration of the particles (see 
also (Ostriker and Cowie 1981)\markcite{OsC}). 
An emphasis was given to these processes in the clusters of galaxies.
In the early works the bright phase was assumed to be 
at relatively large red 
shifts, e.g. $2< z < 10$ in the work by 
Schwartz, Ostriker and Yahil (1975)\markcite{SOY}.

In most recent calculations of Volk et al. (1996)\markcite{Volk} 
the bright phase 
(starburst phase) was placed at relatively low red shift $z \sim 2 - 3$
and the total energy release was found from the iron abundance in the 
clusters. In the discussion below we shall follow this approach.

Finally we  mention the detailed calculations by White and Frenk (1991)
\markcite{WhFr} for the galactic evolution. These calculations start with the 
cosmological evolution of fluctuations described in the Cold Dark Matter 
cosmological model and proceed with the stellar evolution up to the present 
age of the Universe. The star formation in this model occurs late: half of 
the stars are born at the epoch with red shift $z<1$. No bright phase is 
found. This result is not a surprise, because the authors use the mass 
function as it is observed at present time, while the main assumption which 
leads to the bright phase is the strong enhancement of the mass spectrum by  
supermassive short-lived stars.

How much can we increase the calculated diffuse fluxes taking into account 
the bright phase production?

The increase can be estimated as the ratio $W_{CR}^{bp}/W_{CR}^{g}$,
where $W_{CR}^{bp}$ and $W_{CR}^g$ are the CR energy release in a 
cluster during the bright 
phase and due to present-day galaxies, respectively. The value 
$W_{CR}^{bp}$ is given by the product of $N_{SN}$, the total number of 
SN explosions in a cluster during the bright phase, and 
$W_{CR}^{SN}=2.8\times 10^{49}~erg$, the total 
energy of CR produced by one SN. The former can be found, following Volk
et al. (1996)\markcite{Volk}, as $M_{Fe}/0.1 M_{\odot}$, where 
$M_{Fe}$ is the mass of iron in the intracluster gas and 
$0.1 M_{\odot}$ is the
mass of iron produced by one SNII explosion. Since the mass fraction of iron
in a cluster is about $\epsilon_{Fe} \approx 7\times 10^{-4}$ we obtain for our 
representative cluster with the mass of gas 
$M_*^{gas}= 1.8 \times 10^{13}h^{-1}M_{\odot}$:
\be
W_{CR}^{bp}=\epsilon_{Fe} \frac{M^{gas}_*}{0.1 M_{\odot}}W_{CR}^{SN}=
4.4\times 10^{60}~erg.
\ee
The value of $W_{CR}^g$ is 
\be
W_{CR}^g=L_p t_0 N_g \approx 7.7 \times 10^{59}~erg
\ee
and thus the factor of the diffuse flux increase is rather small, about 6.\par
Note that  $W_{CR}^{bp}$  as estimated 
by Volk et al. (1996)\markcite{Volk} 
is $\sim 15$ times larger than our value.
The main reason for this discrepancy is in the value of 
$W_{CR}^{SN}$. Volk et al. adopted it from theoretical considerations 
as $3\times 10^{50}~erg$. Our value is found from phenomenology of CR in 
our Galaxy. Namely it is given as $L_P/\nu_{SN}$, where 
$\nu_{SN} \approx 1/30~yr$ is the frequency of SN explosions in our Galaxy
and $L_P$ is the CR luminosity of our galaxy, given  by 
Drury et al. (1989)\markcite{DMV} as

\be
L_P=cM_g \int dE\frac{\omega_{CR}(E)}{x(E)} \approx 3\times 10^{40}~erg/s,
\ee   
where $\omega_{CR}(E)$ is the observed density of CR in particles with 
kinetic energy $E$, 
$x(E)$ is  the column density  of gas ("grammage") traversed by particles with 
energy $E$ in the 
Galaxy and $M_g~$ is the total mass of the interstellar gas. For a 
reasonable set of these values our present estimates coincide with the 
luminosity given by Drury et al. (1989).

As a result, our value of $W_{CR}^{SN} \approx 2.8\times 10^{49}~erg$ is 11 
times less than that of Volk et al. (1996).\par
We conclude, therefore, that the increase of the diffuse flux due to bright
phase as compared with normal galaxies is smaller than that due to AGN and
accretion shocks.

\section{Discussion and conclusions}
We have calculated the diffuse fluxes of high energy gamma and neutrino 
radiation from clusters of galaxies. These radiations are produced in
pp-interactions of high energy CR with intracluster gas.
The basic assumption which we use is 
that a considerable part of the cosmologically produced 
baryons are concentrated inside
the clusters. An interesting observation is that, with this assumption,
a very restricted set of parameters enters the calculations. They are:
the generation function of CR in the sources, $Q_{CR}(E)$, the cluster size 
$R_{cl}$ and the richness $N_g$. The exact picture of propagation of CR in a 
cluster does not affect the produced flux of radiation, because 
the escape time of CR from a cluster is larger than the age of the cluster.
For the same reason, the spectrum of the produced radiation is very 
flat. At high energy its exponent equals to that of {\em production} spectrum 
of CR in the sources.

We made explicit calculations for CR produced by normal galaxies. In this 
case the generation function, $Q_{CR}(E)$, for a galaxy can be taken as 
that of our Galaxy. It gave us a reference case for calculations: for other 
sources the produced
flux is scaled as the CR luminosity of a source $L_{CR}$.

What other possible sources could there be?

The general restriction is that they must be very powerful to exceed the 
total CR luminosity of normal galaxies in a representative cluster, 
$3\times 10^{42}~erg/s$. 

The anticipated sources could be 
AGN's (including the central cD-galaxy in a cluster) and the shock 
in the accretion flow of the gas onto clusters. They can increase $L_{CR}$
and thus the diffuse flux by factor $\sim 30$. According to our estimates
the bright phase can increase $W_{CR}$, and thus the diffuse flux, only by 
factor 6 as compared with normal galaxies.

The normal galaxies in the cluster can provide neither the observed diffuse 
flux of gamma rays nor the detectable neutrino flux at any energy. The 
accretion shock acceleration and/or acceleration of particles in AGN's
can result in the production of a detectable bump above the atmospheric 
neutrino flux at energy $E \sim 10^6~GeV$, while the 
predicted $0.1 - 100~GeV$
flux of gamma radiation remains one order of magnitude less than observed.

Our results differ from calculations of Dar and Shaviv (1995a, 1995b)
\markcite{DSprl}\markcite{DSaph}: they 
are more pessimistic for gamma radiation and more 
optimistic for neutrino radiation. 

Dar and Shaviv assumed that CR density in clusters is the same as in our 
Galaxy. We cannot reproduce their assumption studying all possible sources
of CR in the cluster. For the CR production by the normal galaxies the 
energy density of CR in a cluster is
$$
\omega_{CR}^{cl}=\frac{3L_PN_gt_0}{4\pi R_{cl}^3}=9.3\times 10^{-4}h^2~eV/cm^3
$$
to be compared with the galactic value $0.5~eV/cm^3$. Here 
$L_P=3\times 10^{40}~erg/s$ is the CR luminosity of our Galaxy and $N_g=100$
is the richness of a typical cluster. 
Increasing the CR luminosity of a cluster up to 
$10^{44}~erg/s$, due to accretion shock or AGN, we are still left with more 
than one order of magnitude deficit.

The bright phase deserves a special discussion. If SNe are the 
main sources of CR, the bright phase gives the increased flux described as
compared to normal galaxies by a factor $n_{SN}^{bp}/n_{SN}^{ng}$,
where $n_{SN}^{bp}$ and $n_{SN}^{ng}$ are the SN space density due to 
bright phase and normal galaxies, respectively. They can be found as
$$
n_{SN}^{bp}= \frac{\epsilon_{Fe}}{0.1 M_{\odot}}
\int_{M_{min}}^{\infty}n(M)M_{gas}dM
$$
and
$$
n_{SN}^{ng}=\nu_{SN} t_0 \int_{M_{min}}^{\infty} n(M) N_g(M) dM
$$
where $n(M)$ is obtained by differentiating eq. (\ref{eq:mf}), $M$ is the
gravitational mass and $M_{gas}\sim 0.1~M$ is the gas mass in a cluster; 
$N_g(M)$ is the number of galaxies in a cluster with mass $M$, and
 $\nu_{SN}$ is the SN frequency in a normal galaxy. 
The ratio $n_{SN}^{bp}/n_{SN}^{ng}$ is about 8, i.e. practically the same as
the previously 
estimated value assuming all clusters having the characteristic mass $M_*$. 

As we demonstrated the spectrum of high energy radiation coincides with the
production spectrum of CR. That is why our spectrum can be as flat as 
$E^{-2.1}$, while Dar and Shaviv explicitly state that their spectrum is
$E^{-2.7}$. This is the reason that our flux of neutrino radiation is 
higher for the same CR luminosity.

The bright phase can give a much more pronounced effect for the case of
{\em individual} cluster.
In particular for the Perseus cluster, discussed 
by Volk et al. (1996), the mass of the 
gas is very large $M \approx 4\times 10^{14} M_{\odot}$ for the given 
richness $N \approx 500$, and therefore the total number of SN is very large 
as compared with the number of SN produced by 
galaxies during their normal 
evolution. Such clusters are the best candidates for observations of both, 
gamma rays and blue-shifted young galaxies within.

Finally, we shall discuss the possibility of gamma-ray detection from 
the nearest Virgo cluster ($19~Mpc$).  

The EGRET observations (Fichtel et al. 1994)\markcite{Fich} give 
the upper limit for $100~MeV$ gamma-ray flux
$F_{\gamma}(>100MeV)<0.5\times 10^{-7}~cm^{-2}~s^{-1}$, which is actually 
valid only for the central source $M 87$. In the models considered here the 
flux from the extended region occupied by the whole cluster can be 
considerably larger. Using the richness N=2500 and mass of the gas 
$M \approx 1\times 10^{14} M_{\odot}$ we obtain the flux at $E_{\gamma} >
100~MeV$  $\sim 10^{-8} cm^{-2} s^{-1}$ both for normal galaxies 
and bright phase. The gain factor predicted for a general case is 
lost for the Virgo
cluster due to the huge number of galaxies in comparison with the gas mass. 
Unfortunatelly this flux is undetectable for the existing instruments.

\section{Acknowledgements}

The authors are grateful to the referee H.J.V\"{o}lk for many useful 
remarks.
\newpage
$ $
\vskip 6cm
\begin{table}[hbt]
\caption {The values of the yields multiplied by $10^3$\label{table_1}}
\begin{center}
\begin{tabular}{ c c c c c c c c }
\hline \hline
$\gamma$ & $2.1$ & $2.2$ & $2.3$ & $2.4$ & $2.5$ & $2.6$ & $2.7$   \\ \hline 
$\gamma-rays$ & $116$ & $88.8$ & $69.0$ & $54.2$ & $43.0$ & $34.5$ & \\   
$\nu_{\mu}+\bar{\nu}_{\mu}$ & $126.2$ & $94.6$ & $69.8$ & $51.9$ & $39.1$ & 
$29.7$ & $22.8$ \\ 
$\nu_{e}+\bar{\nu}_e$ & $58.7$ & $44$ & $32.3$ & $23.81$ & $17.87$ & $13.46$ & 
$10.25$  \\ \hline
\end{tabular}
\end{center}
\end{table}

\newpage

\figcaption[fig1.ps]{The diffuse neutrino flux due to the interactions of the 
CR produced by Normal Galaxies and AGN with the intracluster gas. The upper
limit refers to the maximum neutrino flux correspondent to the observed 
diffuse gamma ray flux as produced in clusters. As a comparison the atmospheric
neutrino flux is plotted with the prompt neutrinos taken into account.
\label{fig_1}}

\end{document}